\documentstyle[12pt]{article}
 \title{REEXAMINATION OF $K\bar{K}$ THRESHOLD  PHENOMENA
 WITH
 $K^+K^-$  ATOM INCLUDED}
 \author{S.V.Bashinsky and B.Kerbikov\\ Institute of Theoretical and
 Experimental Physics\\ Moscow 117259, Russia\\ \\} \date{}

\newcommand{\be}{\begin{equation}}
\newcommand{\ee}{\end{equation}}
\begin{document}
\maketitle

\begin{abstract}
We develop a general framework to study $K\bar{K}$ threshold
phenomena with resonances $f_0$, $a_0$ and
$K^+K^-$ atom included. Based on  this formalism we predict that the
production of the $K^+K^-$ atom in  $pd\to ~^3HeX$ and similar
reactions exhibits a drastic  energy dependence due to the
interplay with  resonances  $f_0$ (980) and
 $a_0$ (980). We
point out that a set of few parameters  describes
 $K\bar{K}$ threshold effects, $f_0$ and $a_0$ mesons, and $K^+K^-$
 atom. Our hope is that precision experimental study aimed at
 determining these parameters may shed more light on the nature of
 $f_0$ and $a_0$ resonances.\\ \\
 PACS number(s): 36.10.Gv, 14.40.Cs, 13.25.-k
   \end{abstract}

\section{Introduction}

It is well known that there is an interesting sometimes controversial
physics bearing upon the $K\bar{K}$ threshold phenomena
 and the nature of $f_0,$ and $a_0$ resonances. It is not the aim of
 our paper to give even a brief  summary on this topic (see, e.g.
Refs.  [1-7]). Our goal is to argue that a  valuable piece of new
information on the subject may come from the study of an exotic
$K^+K^-$ atom (kaonium).
The main point of this paper is that kaonium might be included into
the effective Hamiltonian which describes $K\bar{K}$ threshold effects
and resonances $f_0, a_0$. Such a special approach to kaonium as
compared to other exotic atoms is due to the fact that $K\bar{K}$
threshold region is completely overlapped by resonances $f_0$ and
$a_0$ to both of which kaonium is coupled. Therefore examining
kaonium we face the interference of several overlapping resonances.
It is well known that such interference gives rise to various
intriguing phenomena. In our case this will be the drastic energy
behaviour of kaonium production amplitude. Our interest in describing
kaonium furnished a major motivation of constructing the
phenomenological effective Hamiltonian for the $K\bar{K}$ threshold
region. Having constructed such a Hamiltonian we find ourself in a
position to study $K\bar{K}$ threshold anomalies and the mixing of
$f_0$ and $a_0$ mesons due to the $K^0\bar{K}^0$ and $K^+K^-$
thresholds splitting.

The organization of the paper is the following. The rest of this
Section is devoted to introductory remarks on kaonium. Then in
Section 2 we consider the specific reaction $pd\to ~^3HeX$ in which
kaonium may be possibly observed. Our main formalism is presented in
Section 3 where the effective Hamiltonian is constructed.
In Section 4 we study the production of kaonium in its ground state
and predict a drastic energy dependence of this process mentioned
above. Section 5 is devoted to $K\bar{K}$ threshold anomalies.
Section 6 contains the discussion about the mixing  of $f_0$ and
$a_0$ mesons. In Section 7 we present the set of parameters behind
the effective Hamiltonian, and indicate those which have to be
determined from future experiments. In Section 8 we estimate the
mixing parameters of kaonium and resonances $f_0,~ a_0$ making use of
the alternative models for the $f_0$ and $a_0$ structure. Our main
results are summarized in Conclusions.

Kaonium has not been observed yet but corresponding experiments are
contemplated --see the paper [8] and references therein.
To our knowledge the kaonium production was first discussed by Wycech
and Green [8]. We have already mentioned that our study of
kaonium is based on the  idea of the interplay of kaonium and
resonances  $f_0$   and $a_0$. This makes our approach
 (first outlined in [9]) different from that of
[8]. We are not aware of any  previous discussion of the mixing of an
exotic atom and a hadron resonance.

Kaonium has the  Bohr radius of $a_B=2/\alpha m_{K^+}=109.6 fm$, the
ground state binding energy is $E_0=\alpha^2m_{K^+}/4=6.57 keV$. The
expected decay width is $\Gamma=\alpha^3m^2_{K^+} Im A\simeq 640 eV $
[8], an alternative estimate for $ \Gamma$ will be given below.
Kaonium is formed by electromagnetic interaction and its wave
function has two isospin components with $I=0$ and $I=1$.  Therefore
it is mixed with both  $f_0$ and $a_0$; different facets of this
mixing will be discussed at length in what follows.

\section{ $pd\to~^3HeX$ reaction close to $K\bar{K}$ thresholds}

To be specific we shall consider $f_0,a_0$ meson and  kaonium
production in the reaction
\be
 pd\to ~^3HeX\to ~^3He\pi^+\pi^-
 \ee
  with
the invariant mass  of $X$ being in the vicinity of $K^+K^-$ and
$K^0\bar{K}^0$ thresholds.  The $K^+K^-$ production threshold in
reaction (1) corresponds to the kinetic energy of the incident proton
beam equal to $T_p=1.73 GeV$. A precision experimental study of the
$K\bar{K}$ threshold region in the reaction (1) is a part of
the program at SATURNE [10] and  at COSY [11]. Although we shall
discuss  kaonium production only in the reaction (1), the formalism to
be developed can be easily applied to kaonium production in other
processes as well.

At this point a remark is in order. The term kaonium production used
above is somewhat misleading. In what follows we shall explain in
detail that the object $X$ produced in reaction (1) close to $K^+K^-$
threshold is a mixture of $f_0$, $a_0$ mesons and kaonium atom.
Therefore the question of what is the kaonium production probability
in reaction (1) has to be  replaced by the question of what is
the admixture of kaonium in $X$ and how this admixture manifests
itself experimentally. The answer to the last question is one of the
 purposes of our paper.

We start with a standard equation for the cross section  of reaction
(1)
\be
d\sigma= (2\pi)^4\delta^{(4)}(P_p+P_d-P_{He}-P_{\pi^+}-P_{\pi^-})
\frac{|{\cal M}|^2}{4I}\prod_f\frac{d^3{\bf p}_f}{(2\pi)^32E_f},
\ee
where the product runs over $^3He,~\pi^+$ and $\pi^-$. In a typical
experimental situation one is interested in the double differential
cross section $d^2\sigma/dp_{He}d\Omega_{He}$ (here $p_{He}$ stands
for $|{\bf p}_{He}|$) corresponding to the measurement of helium
momentum and direction in the laboratory frame. Routine arguments
allow to perform phase space integration with $|{\cal M}|^2$
factorized out. The result reads
\be
\frac{d^2\sigma}{dp_{He}d\Omega_{He}}=| {\cal
M}|^2\frac{p^2_{He}}{2^8\pi^4m_X^2E_{He}}
\frac{\lambda^{1/2}(m^2_X;m^2_{\pi},m^2_{\pi})}{\lambda^{1/2}(s;m^2_p,m^2_d)},
\ee
where $s=(P_p+P_d)^2,~~m_X$  is the invariant mass of $X$ and
$\lambda(x;y,z)=(x-y-z)^2-4yz $
 is a kinematical function.
 In what follows use will be made of
the $T$--matrix connected to the invariant amplitude
 via the relation
  \be
   T={\cal M}
\prod_{i,f}(2E_i)^{-1/2}(2E_f)^{-1/2},
 \ee
  where the initial state
index $i$ includes $p$ and $d$, and the final state index $f$ runs
over $^3He$, $\pi^+$, $\pi^-$.
It is simply related to  $S$--
matrices as
\be
 S=1+i(2\pi)^4\delta^{(4)}(P_f-P_i)T,
  \ee
where $P_f$ and $P_i$ are the sums of the final and
initial particles 4--momenta.
   In the dynamical model to follow the $T$--matrix will be
considered as a function of the invariant mass $m_X$. Therefore
Eqs.(3-4) should be supplemented by a kinematical relation between
laboratory frame momentum of $^3He$ and the invariant mass $m_X$. One
defines $\theta$ as the angle between the direction of the
incident proton and outgoing $^3He$. Then the following relation
is obtained
\be
m_X^2=m^2_p+m^2_d+m^2_{He}+2E_p(m_d-E_{He})-2m_dE_{He}+2P_pP_{He}cos\theta.
\ee
According to (6) one can vary $m_X$ either by changing the
momentum of the incident proton with $^3He$ momentum and angle
kept fixed, or by measuring $^3He$ spectrum with initial proton
momentum  kept fixed.

Our next goal is to get an expression for the
$T$-- matrix which corresponds to our physical picture in which $f_0$
and $a_0$ mesons dominate the $K\bar{K}$ threshold region and hence
the kaonium production.

\section{The $T$--matrix and the effective Hamiltonian}

The literature on overlapping resonances, resonance "mixtures", and
related topics is overwhelming. The formalism to be used  here is
based  on two essentially equivalent approaches. The first one is the
mass-matrix method which is very transparently presented in a paper
by Kobzarev, Nikolaev and Okun [12]. An alternative way to get
similar results was developed by Stodolsky [13] in what he called
scattering theory of resonance "mixtures". We find  it hardly
necessary to retrieve here the basic equations. Rather we would like
to concentrate on novel features which characterize the mixing of an
atom and a resonance.

Consider reaction (1) with $X$ being a mixture of $f_0$, $a_0$, and
kaonium. The  detailed composition of $X$ will become clear below. Here we
only mention that the mixing of $f_0$ and $a_0$ is possible via the coupling
of both to kaonium. Then $T$ -- matrix, defined by Eqs.(4) and (5),
satisfies the following equation
\be
 T=<\pi\pi|\hat{V}\hat{G}|A>+T^0
.  \ee

First we explain the notations in (7) and then consider each of its
blocks in detail. The $|\pi\pi>$ stands for the final $^3He\pi\pi$
system with definite momenta, $\hat{V}$ is the transition operator
corresponding to the $f_0\to \pi\pi$ process (we remind that the
decay $a_0\to\pi\pi$ is forbidden). The Green's function $\hat{G}$
describes $f_0$--$a_0$ system plus $^3He$ spectator  and its coupling
 to resonance decay channels as well as to the kaonium.  $|A>$ is a
 (two component) creation amplitude of resonances $f_0$, $a_0$ plus
$^3He$ spectator.  Into the background term $T^0$ we have lumped
direct $^3He\pi\pi$ production.

Now we consider the Green's function $\hat{G}$ of mesons $f_0$
and $a_0$ ($^3He$ plays the role of spectator, i.e.  it enters only
through kinematics).
 We shall write $\hat{G}$ as a $2\times 2$ matrix in the
basis corresponding to "bare" $f_0$ and $a_0$, which we denote as
$|f>$ and $|a>$. The term "bare" means that $|f>$ and $|a>$ are not
mixed with decay channels and kaonium.
 These states are assumed to have
definite isospin $I=0$ for $|f>$ and $I=1$ for $|a>$.
Since we consider reaction (1) in the vicinity of $K\bar{K}$
thresholds the relative motion in the system $K\bar{K}$ can be treated
nonrelativistically, which will be used in
writing expressions for the corresponding propagators . All the
above remarks on the operator $\hat{G}$ allow to write it in the form
 \be
  \hat{G} = \left( \begin{array}{cc} G_{ff}&G_{fa}\\
G_{af}&G_{aa} \end{array} \right)
=(m_X-\hat{H}-\Delta\hat{H})^{-1},
\ee
 where $m_X$ is the
invariant mass of the system $X$ in reaction (1), $\hat{H}$
stands for the Hamiltonian  of the bare $|f>$ and $|a>$
\be
\hat{H}=\left( \begin{array} {ll} E_f^{(0)}&0\\ 0&E_a^{(0)}
\end{array}  \right),
\ee
 while their coupling to decay channels
$\pi\pi,~\pi\eta,~ K^+K^-,K^0\bar{K}^0$  and kaonium is described
 by the matrix $ \Delta\hat{H}$. It has the form
  \be
\Delta\hat{H}=\sum_{\alpha}\hat{V}^
+\frac{|\alpha><\alpha|}{m_X-m_{\alpha}+i0} \hat{V},
\ee
where $|\alpha>$ indicates any of the states ($\pi\pi,~\pi\eta,
K^+K^-,~K^0\bar{K}^0$, kaonium) to which $|f>$ and $|a>$ are
coupled, and $m_{\alpha}$ is its invariant mass. In the basis
employed in Eq.(8) the transition operator $\hat{V}$ is represented
by a row $(V_f,V_a)$, while the $f_0$ and $a_0$  production amplitude
$A$ by a column $\left( \begin{array} {c} A_f\\ A_a\end{array}
\right)$.  Now we must be reminded of the fact that due to the
$G$--parity conservation only $V_f$ element has nonvanishing
projection on the final $^3~He\pi\pi$ state. Therefore inserting Eq.
(8) into (7), we get the following expression for the amplitude
$T$:  \be T=V_{\pi\pi,f}G_{ff}A_f+V_{\pi\pi,f}G_{fa}A_a+T^0, \ee
where
\be V_{\pi\pi,f}\equiv <\pi\pi|\hat{V}|f>. \ee

Equation (11) represents the gist  of our approach.
Now let us consider the contribution of the kaonium ground  state
into the $T$--matrix (11).

\section{The amplitude behaviour near the kaonium ground state}

In this section we shall discuss how the kaonium
ground state  manifests itself in the $T$--matrix (11) and in the
cross section (3). To this end
                                in Eq.
(10) for $\Delta\hat{H}$ we single out from the sum the term
corresponding to kaonium ground state, while the remainder add to
$\hat{H}$  given by Eq.  (9).  Such a regrouping yields \be
\hat{H}+\Delta\hat{H}
=\frac{\hat{V}^+|at><at|\hat{V}}{m_X-m_{at}}+ \left( \begin{array}{ll}
E_f-i\frac{\Gamma_f}{2}&0\\ 0&E_a-i\frac{\Gamma_a}{2} \end{array} \right),
\ee
 where the state $|at>$ is the kaonium ground state with a pure Coulomb
binding energy, i.e. $m_{at}=2m_K-6.57 keV$. In the next Section we
shall see that the second term of Eq. (13) displays only minor
 energy dependence within the $m_X$ interval smaller than the
 distance between the  ground and first excited states of kaonium.
 Therefore in this Section the quantities $E_f, E_a, \Gamma_f,
 \Gamma_a$ are considered as energy independent. The corrections
to this approximation are essential for the cusp structure at
$K^+K^-$ threshold which will be considered in the next Section.
  We have  also neglect the nondiagonal
elements in the second term of (13).
This amounts to neglecting the
$f_0-a_0$ mixing via kaonium excited states and, what is much more
substantial, via the low  energy parts of the  $K^+K^-$ and
$K^0\bar{K^0}$ continuous spectra.  The validity of
such approximation will be discussed in the Section 6.

Now we substitute the effective Hamiltonian (13) into the Green's
function (8) and $T$--matrix (11). After a  few formal manipulations
this yields
$$
T=\frac{V_{\pi\pi,f}A_f}{m_X-E_f+i\frac{\Gamma_f}{2}}+
$$
\be
+\frac{1}{(m_X-E_{at}+i\frac{\Gamma_{at}}{2})}
\frac{V_{\pi\pi,f}V^+_{f,at}}{(m_{at}-E_f+i\frac{\Gamma_f}{2})}
\left[\frac{V_{at,f}A_f}{m_{at}-E_f+i\frac{\Gamma_f}{2}}+\right.
\ee
$$
\left. +\frac{V_{at,a}A_a}{m_{at}-E_a+i\frac{\Gamma_a}{2}}
\right]+T^0, $$
where $V_{at,r}=<at|\hat{V}|r>,~~r=f,a,$ and
 \be
E_{at}=m_{at}+\sum_{r=f,a}|V_{at,r}|^2\frac{m_{at}-E_r}{(E_r-m_{at})^2
+\frac{\Gamma_r^2}{4}},
\ee
\be
\Gamma_{at}=\sum_{r=f,a}|V_{at,r}|^2\frac{\Gamma_r}{(E_r-m_{at})^2
+\frac{\Gamma_r^2}{4}}.
\ee

Formulae (14-16) are the sought--for equations which display the
effect of the kaonium ground state in reaction (1). They are
reminiscent of equations describing giant nuclear resonances (see,
e.g. [14-15]).

Equation (14) shows that as a function of  the invariant mass $m_X$,
the amplitude $T$ in the reaction (1) has the pole  at
$m_X=E_{at}-i \Gamma_{at}/2$. It corresponds to the
atomic level which has acquired the shift (15) and the width (16) due
to the mixing of $f_0$ and $a_0$ mesons.

The first term in Eq.(14) looks  as if
$T(m_X)$ had another pole at $m_X=E_f-i\Gamma_f/2$ corresponding to
 the $f_0$ resonance.  The issue, however, is more subtle.
 The simple pole structure  for $f_0$ is invalidated by the energy
 dependence of its width  $\Gamma_f$ and the existence of the
 so--called "shadow" poles on different sheets of the energy
 Rieman surface [3,4,16]. We shall return to this point in Section
 6, for more discussion see the cited references.

Equation (14) may be presented in a
more concise and transparent form by introducing the following "propagators"
\be
g_f(m)=\frac{1}{m-E_f+i\frac{\Gamma_f}{2}},~~
g_a(m)=\frac{1}{m-E_a+i\frac{\Gamma_a}{2}},~~
g_{at}(m)=\frac{1}{m-E_{at}+i\frac{\Gamma_{at}}{2}}
\ee
Then (14) takes the form
\be
T=V_{\pi\pi,f}g_f(m_X)A_f+
V_{\pi\pi,f}g_f(m_{at})V^+_{f,at}g_{at}(m_X)
[V_{at,f}g_f(m_{at})A_f+V_{at,a}g_a(m_{at})A_a]+T^0.
\ee
The first and the second terms of Eq.(18) may be represented by
diagrams in Figs. 1(a) and 1(b) correspondingly.

 In order to display
the behaviour of the cross section (3)
in the vicinity of the kaonium ground state we introduce the quantities

\be \Delta E_{at}^{(r)}=
V_{at,r}g_r(m_{at})V^+_{r,at}=
\frac{|V_{at,r}|^2}{m_{at}-E_r+i\frac{\Gamma_r}{2}},~~~~r=f,a.  \ee
which have a transparent interpretation of the complex energy
shift of the atomic level due to its mixing with a resonance $r$. In
fact, Eqs. (15-16) yield
\be
E_{at}-i\frac{\Gamma_{at}}{2}= m_{at }+\Delta E_{at}^{(f)}+\Delta
E_{at}^{(a)}.
\ee
In these notations the amplitude $T$ (18) reads
\be
T=V_{\pi\pi,f}g_f(m_{at}) \{ 1 + \frac{\Delta
E_{at}^{(f)}+\eta\Delta E_{at}^{(a)}}{m_X-m_{at}-\Delta
E_{at}^{(f)}-\Delta E_{at}^{(a)}}\}A_f+T^0,
 \ee
where
\be
\eta=\frac{V^+_{f,at}A_a}{V^+_{a,at}A_f}.
\ee
 A careful look at Eqs. (21-22) leads to striking observation.
 Suppose for a moment that the background term $T^0$ as well as the
 coupling to $a_0$ meson are absent $(T^0=0,~\Delta E^{(a)}_{at}=0).$
 Then the amplitude $T$ has a zero at $m_X=m_{at}$ (compare to the
 well--known "dipole" phenomena considered in chapter 8 of [17]).
  The
 structure above is similar to what is called "Fano zero" [18]
 of the amplitude.

 The general case corresponds to a finite  value
 of the background term $T^0$ and  a certain (complex) value of the
 parameter $\eta$. Unless the background becomes dominant, the cross
 section still has a dip -- bump structure at the vicinity of the
 atomic level.
 The exact position of the minimum as well as the height of the
 adjacent peak depends upon the parameter $\eta$. In particular from
 Eq.(21) one notices that for $\eta=1$ the first term of (21) again
 has a zero at the same point $m_X=m_{at}$.

   From Eqs. (19), (21) it follows that
 the energy scale of the zero--peak (or dip--bump) structure is
 governed by the mixing parameter $|V_{r,at}|^2,~~r=f,a$. This
 quantity, which is sensitive to the  nature of $f_0$, $a_0$ mesons,
 is discussed in Sections 7, 8
 .

 In Fig. 2 we plot the typical cross section $
 d^2\sigma/dp_{He}d\Omega_{He}$ (see Eq. (3)) as  a function of
 $m_X$.  We remind that $m_X$ is connected to $^3He$ momentum and
 angle via relation (6). The set of parameters behind Fig. 2 is
  discussed in Section 7.

  \section{$K\bar{K}$ threshold effects}

  In this section we give an account of anomalies which show up
  at $K\bar{K}$ thresholds. The $K^+K^-$ and $K^0\bar{K}^0$
  thresholds are  splitted by 8 MeV, and due
  to the Coulomb attraction in the $K^+K^-$ channel the cusp
    phenomena are quite different. First we consider the case
    of charged kaons.

  The  general character of threshold anomaly in the case of Coulomb
  attraction is well known. As the energy of the $\pi^+\pi^-$ system
  in reaction (1) approaches the $K^+K^-$ threshold from below, we
  will observe  a series of  ever more rapid oscillations due to the
  excitation of the infinitely numerous Coulomb levels of kaonium.
  The limit of the cross section (3) below threshold does not exist
  since the threshold energy is the accumulation point for these
  resonances.
   The purpose of this section is to reproduce these general results
   from our equations and to investigate which of the parameters of
   our model may be determined from precision measurements of the
   $K^+K^-$ threshold effects.

   Again we start with regrouping different terms of the effective
   Hamiltonian $\hat{H}+\Delta\hat{H}$ given by Eqs. (9-10).
   This time we single out from $\Delta\hat{H}$ (10) all
   contributions stemming from $S$--wave states of the $K^+K^-$
   system (atomic levels plus continuous spectrum).  The  transition
   operator $\hat{V}$ (see Eq. (10) and the text afterwards) is of
   short--range character and hence dominated by high momentum
   components. The $K^+K^- $ system near threshold is on the contrary
   characterized by low momenta. Therefore one has \be
   \hat{V}^+|\alpha>= \int
   \frac{d^3p}{(2\pi)^3}\hat{V}^+|\vec{p}><\vec{p}|\alpha>\simeq
   \hat{V}^+|K^+K^->\psi_{\alpha}(0),
   \ee
   where use was made of the fact that the main contribution to
   the integral comes from small momenta typical for
   $<\vec{p}|\alpha>$, at these momenta $\hat{V}^+|\vec{p}>$ can be taken
   out of the integral as a momentum independent vector
   $\hat{V}^+|K^+K^->$, and the remaining integral yields the
   $K^+K^-$ system wave function at the origin $\psi_{\alpha}(0)$.

   In particular, the parameter $V_{at,r}=<at|\hat{V}|r>~~(r=f,a)$,
   which characterizes the mixing of kaonium ground state with the
   corresponding mesons, takes the form
   \be V_{at,r}=\psi_{at}(0)<K^+K^-|\hat{V}|r>=
   \sqrt{\frac{\alpha^3m^3_K}{8\pi}}<K^+K^-|\hat{V}|r>
   \ee
   The geometrical factor has been singled out,
   and the remaining "reduced" matrix element
   $<K^+K^-|V|r>$ depends on the nature of $f_0$, $a_0$ mesons. It
   will be discussed in Section 8.

    Summation in Eq.(10) over $K^+K^-$ states implies both
   integration over continuous spectrum and summation over atomic
   levels.  Therefore we have \be
   \Delta\hat{H}=\Delta\hat{H'}+\hat{V}^+|K^+K^-><K^+K^-|\hat{V}
   \left[
   \int\frac{d^3p}{(2\pi)^3}\frac{|\psi_p(0)|^2}{\Delta
   m-\frac{p^2}{m_{K^+}}+i0}+\sum^{\infty}_{n=1}
   \frac{|\psi_n(0)|^2}{\Delta m+\frac{m_K\alpha^2}{4n^2}}\right],
   \ee
   where $\Delta m = m_{X}-2 m_{K^+}$, and
   \be
   |\psi_n(0)|^2=\frac{\alpha^3m^3_K}{8\pi n^3},~~
   |\psi_p(0)|^2=\frac{\pi\alpha m_K}{p\{
   1-exp(-\frac{\pi\alpha m_K}{p})\}}.
   \ee
   The remaining part $\Delta\hat{H}'$ in Eq.(25) arises from the
    $f_0$ and $a_0$ decay channels
   other than $K^+K^-$ (i.e. $\pi\pi,\pi\eta,K^0\bar{K}^0$).

       The integral
   in (25) diverges at large values of $p$.  This is a formal
   difficulty since at large momenta the operator $\hat{V}$ induces
   cutoff (also at large values of $p$ one has to use relativistic
   kinematics).  Within the energy interval $|\Delta
   m|=|m_X-2m_{K^+}|\leq E_0=\frac{\alpha^2m_{K^+}}{4}$,
   where
   Coulomb effects are important, Eq.(25) yields
   \be \Delta
   \hat{H}=\Delta\hat{H}'+\hat{V}^+|K^+K^-><K^+K^-|\hat{V} \left[\Phi
   +\frac{\alpha m^2_{K^+}}{4}\left \{
   \begin{array}{ll}
   cot(\pi\sqrt{\frac{m_{K^+}\alpha^2}{4|\Delta m|}}),& \Delta m<0\\
   -i,&\Delta m>0 \end{array}       \right.
   \right],
   \ee
   where $\Phi$  is a real function which experiences smooth
   variations not exceeding $\alpha m^2_K$ over the interval $|\Delta
    m|\leq E_0$.

    The Equations (8),(11),(27) retrieve  the Coulomb threshold
    anomaly as it  was outlined at  the beginning of this Section.
   Eq.(27)  also proves  the validity of the approximation (13) with
   constant $E_r$ and $\Gamma_r(r=f,a)$ at the immediate vicinity of
   the  kaonium ground state.

   Now we briefly discuss what happens at the $K^0\bar{K}^0$
   threshold. Here the discrete spectrum  is absent, and instead of
   Eq. (26) we have $|\psi_p(0)|^2=1$. Equation (25) takes the form
   \be
   \Delta\hat{H}
   =\Delta\hat{H}^{''}+\hat{V}^+|K^0\bar{K}^0><K^0\bar{K}^0|\hat{V}\int
   \frac{d^3p}{(2\pi)^3}\frac{1}{\Delta m-\frac{p^2}{m_{K^0}}+i0},
   \ee
   with $\Delta m=m_X-2m_{K^0}$ and the same remarks
   applied  concerning the
   convergency of the integral.  The energy dependent part of (28)
   reads
   \be
   \Delta\hat{H}^{(n)}=\hat{V}^+|K^0\bar{K}^0><K^0\bar{K}^0|\hat{V}\left\{
   \begin{array}
   {ll}
   \frac{m^2_{K^0}}{4\pi}\sqrt{\frac{|\Delta m|}{m_{K^0}}},& \Delta
   m<0\\
   -i \frac{m^2_{K^0}}{4\pi}\sqrt{\frac{|\Delta m|}{m_{K^0}}},&
   \Delta m>0.
   \end{array}
   \right.
   \ee
   Thus at $K^0\bar{K}^0$ threshold the amplitude  displays the
   standard  behaviour of the  $\sqrt{2m_{K^0}-m_{X}}$ type. The same
   pattern also restores in the $K^+K^-$ channel when
   $|m_X-2m_{K^+}|\gg E_0$, i.e. away of the narrow Coulomb
   region.
     Eq. (29) and its analogue for charged kaons will be the starting
   points of the next Section.

    \section{Mixing of the  $f_0$ and $a_0$ mesons}

      Now we proceed to the energy scale of the order of the interval
      between $K^+K^-$ and $K^0\bar{K}^0$ thresholds:
$2m_{K^0}-2m_{K^+}=8MeV$. It by far exceeds a characteristic Coulomb
energy $E_0=6.6 keV$ therefore we disregard Coulomb effects in the
present section. Then the energy dependence of the effective
Hamiltonian $\hat{H}+\Delta\hat{H}$ (9-10) is given by the two
terms of the form (29) corresponding to the charged and neutral kaon
channels
\be
 \hat{H}+\Delta\hat{H}=\hat{H}^{(0)}
+\Delta\hat{H}^{(ch)}(m_X)+\Delta
\hat{H}^{(n)}(m_X)=
\ee
$$
=\hat{H}^{(0)}+\hat{V}^+|K^+K^-><K^+K^-|\hat{V}(-i\frac{m^2_K}{4\pi})\sqrt{
\frac{m_X-2m_{K^+}}{m_K} + i0}~~ +
$$
$$
+\hat{V}^+|K^0 \bar{K}^0><K^0
\bar{K}^0|\hat{V}(-i\frac{m^2_K}{4\pi})\sqrt{
\frac{m_X-2m_{K^0}}{m_K} + i0},
$$

where  adding  $(+i0)$ means that for  $m_X<2m_K$ the square roots
acquire positive imaginary parts, and  where the $K^+$ - $K^0$ mass
difference is ignored except for the under square root expressions.
The first term $\hat{H}^{(0)}$ of Eq.(30)
includes the Hamiltonian $\hat{H}$ (7) of the bare
$|f>$ and $|a>$ states (see Eq.(9)) plus contributions from
$f_0,~~a_0$ decay
channels other than $K\bar{K}$
( such as $\pi\pi$ and $\pi\eta)$.

We now remind that the states $|f>$ and $|a>$, which were introduced in
the Sec. 3, have  definite isospins $I_f=0,~ I_a=1$.  Then isospin
invariance implies
\be
<K^+K^-|\hat{V}|f> =<K^0\bar{K}^0|\hat{V}|f>,~~
<K^+K^-|\hat{V}|a> =-<K^0\bar{K}^0|\hat{V}|a>.
\ee
Having got  these relationships one can put the Eq. (30) into a more
instructive form. For this purpose we introduce the notations
 \be
D=|<K^+K^-|\hat{V}|f>|^2\frac{m^2_K}{4\pi},
\ee
\be
\zeta=\frac{<K^+K^-|\hat{V}|a>}{<K^+K^-|\hat{V}|f>}.
\ee

Then
\be
\hat{H}+\Delta\hat{H}=\left(
\begin{array}{ll}
E'_f-i\Gamma_f/2&0\\
0& E'_a-i\Gamma_a/2  \end{array} \right)+
\ee
$$
+D\left(
\begin{array}{ll}
1&\zeta\\
\bar{\zeta}&|\zeta|^2\end{array} \right)
(-i)\sqrt{\frac{m_X-2m_{K^+}}{m_K}+i0}
+D\left(
\begin{array}{ll}
1&-\zeta\\
-\bar{\zeta}&|\zeta|^2\end{array} \right)
(-i)\sqrt{\frac{m_X-2m_{K^0}}{m_K}+i0}.
$$

In this equation the  first term  is just a
parameterization for $\hat{H}^{(0)}$ in Eq.(30). It must be
diagonal for all its contributions, discussed after  Eq.(30), have
only diagonal elements.

 From Eqs. (34),(8) and (11) it is clear that in the $K\bar{K}$
threshold region the cross section (3) (again considered as a
function of $m_X$) is influenced by both mesons $f_0$ and $a_0$
and also depends on the parameter $\zeta$. Some
typical plots are presented in Fig. 3.

 Far away of the
$K\bar{K}$ thresholds the nondiagonal elements in the second and
third terms on the right hand side of Eq.(34) cancel each other.
However in the threshold region they are quite substantial.  In fact,
if $m_X$ is fixed exactly at the $K^+K^-$ threshold  so that the
second term in (34)  vanishes,  the nondiagonal contribution due to
the third term  amounts to $-\zeta
D\sqrt{2(m_{K^0}-m_{K^+})/m_K}=-27\zeta~ MeV$ which might be compared
to $\Gamma_f/2=108 MeV$ ( numerical estimates for $D$ and
$\Gamma_f$ follow Ref. [4], see Section 7  for more
details).

At this point we return to Eq.(13) describing the effective
    Hamiltonian in the vicinity of kaonium ground state and
   address the question of its validity. To this end we make use of
   (34) and introduce into the second term of (13) the contributions
   hitherto omitted. To be more explicit we put all quantities in MeV
   units. Then the second term of (13) takes the form
    \be \left( \begin{array}{ll}
   E'_f-i\Gamma_f/2+27&-27\zeta\\
   -27\bar{\zeta}&E'_a-i\Gamma_a/2+27|\zeta|^2
   \end{array} \right).
   \ee
   Therefore our approximation (13) with vanishing nondiagonal
   elements in the second term is  appropriate provided
    $f_0$  $K\bar{K}$  coupling constant is not smaller
   than the coupling constant
    $a_0$ $K\bar{K}$. To our knowledge this question is still
   ambiguous and controversial [16].

   Introduction of nondiagonal elements in the last term of Eq.(13)
   does not change the general conclusions of Sec.4  concerning a
   narrow resonance at the kaonium level but corresponding formulas
   will be well complicated. For the complex energy shift of the
   kaonium ground level (20), the expansion in a power series of the
   parameter $\zeta$ gives the following linear term
   \be
   -2\frac{Re(\zeta V^*_{at,a}V_{at,f})}{(m_{at}-E_f+i\Gamma_f/2)
   (m_{at}-E_a+i\Gamma_a/2)}\cdot 27 MeV=
   \ee
   $$
   -\frac{|\psi_{at}(0)|^2|<K^+K^-|\hat{V}|a>|^2
   |<K^+K^-|\hat{V}|f>|^2} {2\pi(m_{at}-E_f+i\Gamma_f/2)
   (m_{at}-E_a+i\Gamma_a/2)}\sqrt{2m^3_K(m_{K^0}-m_{K^+})}
   $$

 \section{Parameters of the model}

In our description of $f_0$ and $a_0$  mesons  we have
 the following set of parameters to be determined:
five real quantities $E'_f, E'_a, \Gamma_f,\Gamma_a,D$ and a complex
quantity $\zeta$; they are  presented by Eqs. (32-34). In addition  we
have two production amplitudes $A_f,A_a$  (see Eq.(11)) and the
transition matrix element $V_{\pi\pi,f}$ (see (11-12)).

The other quantities in our formalism are related to those nine.
Namely, $<K\bar{K}|\hat{V}|r>~~(K=K^+,K^0; ~~r=f,a)$ can be
determined from Eqs. (32-33)while its bearing on the nature of
$f_0,~~a_0$ mesons will be the subject of the next Section.
Parameters $E_f$ and
$E_a$ in Eq. (13) are connected to $E'_f$ and $E'_a$ via Eqs.
(13),(34) as
\be
 E_f=E'_f+D\sqrt{\frac{2m_{K^0}-2m_{K^+}}{m_K}} \ee
$$ E_a=E'_a+|\zeta|^2 D\sqrt{\frac{2m_{K^0}-2m_{K^+}}{m_K}} $$

To obtain the values of $E'_f,~\Gamma_f$ and $D$ we remind that
for $m_X$ well outside the  8 MeV  $K^+K^-$ , $K^0\bar{K}^0$
interthreshold region  the off--diagonal elements of the
effective Hamiltonian $\hat{H}+\Delta\hat{H}$ (34) are cancelled, so
that  $f_0$ and $a_0$  are not mixed.  Then $\pi\pi$
production in the reaction (1) goes only through
$f_0$ meson $(a_0\to\pi\pi$ is  $G$-- parity forbidden).
 Thus the Green's function $\hat{G}$ (8) becomes diagonal
and the amplitude (11) takes the simple form
 \be
T=\frac{V_{\pi\pi,f}A_f}{m_X-E'_f+i\Gamma_f/2+2iD
\sqrt{\frac{m_X-2m_K}{m_K}+i0}}
+T^0,
\ee
which is just a Breit--Wigner amplitude with mass--dependent width.
Recent analysis of Breit--Wigner contribution to the $f_0$ structure
is presented in Refs. [3,4]. In our present paper we stick to the set
of parameters proposed by Zou and Bugg [4]; our set of parameters is
related to their via
$$
E'_f=\frac{M^2_R}{2m_K}=916 MeV $$
\be
\Gamma_f=\frac{g_1}{m_K}\sqrt{1-\frac{m^2_{\pi}}{m_K^2}}=214 MeV
\ee
$$
D=\frac{g_2}{4m_K}=213 MeV
$$
where $M_R,g_1,g_2$ are quantities used in [4]. This yields $E_f=943$
MeV in Eq. (A1) and $\Delta E_{at}^{(f)}=66-162i$ eV in Eq. (17).

As far as the parameters $E'_a,~\Gamma_a$ and $\zeta$ are concerned,
we are not aware of the $a_0$(980) pole structure analysis which
could provide their values.

\section{ Mixing parameter  as a probe to the  nature of
   the $f_0,a_0$ mesons}

   In Section 4 we have shown that the  energy  dependence of the
   zero-peak (dip--bump) structure due to kaonium is governed by the
   quantities $|V_{at,r}|^2,~~r=f,a$. This is where the nature of the
   mesons comes into play. The factorization (24)
   \be
   |V_{at,r}|^2=|\psi_{at}(0)|^2|<K^+K^-|V|r>|^2
   \ee
   shows that the quantity $|V_{at,r}|^2$
   is proportional to the absolute value  square of the
   "reduced" mixing parameter ~~~$<K^+K^-|V|r>$ . The
    geometrical factor $|\psi_{at}(0)|^2$ characterizes an
   overlap of the wave functions.

   In order to be transparent and avoid cumbersome equations, we
   consider a toy model with only one meson, say $f_0$, and try to
   estimate the quantity $|V_{at,f}|^2$ for different hypotheses on
   the nature of this meson [1-5].

   Consider first an interpretation of the $f_0$ as a $K\bar{K}$
   molecule, i.e. a deuteron--like state. A simple estimate
    yields
   \be
   |<K^+K^-|V|f_0>|^2\simeq\frac{8\pi}{m_K}
   (\frac{|\varepsilon-2m_K|}{m_K})^{1/2}=1.1\cdot 10^{-2}MeV^{-1}
   \ee
   where $\varepsilon=988-23i~  MeV$ is the second--sheet pole
   for $f_0$(980) [4].
   This estimate corresponds to numerical value $\Gamma_{at}\simeq 330
   eV$ in Eq.(16).

   For the alternative compositions  of $f_0$ (e.g. $q\bar{q}$,
   multiquark, glueball, etc.), the  natural way to parameterize
   $<K^+K^-|V|f_0>$ is through the Jaffe--Low $P$--matrix [19].

   In $P$--matrix terminology $f_0$ is the $P$--matrix "primitive"
   with eigenvalue $E_n$, radius $b$ and the coupling to hadronic
   channels given by a residue  $\lambda_n$. In terms of these
   quantities the $P$--matrix reads:
   \be
   P=k~cot(kb+\delta)=P_0+\frac{\lambda_n}{E-E_n}
   \ee
   Straightforward but somewhat lengthy calculations [20,21] lead to
   the estimate
   \be
   |<K^+K^-|V|f_0>|^2=\frac{8\pi}{m_k} \lambda_n b^2
   \ee
   The value of the residue $\lambda_n$ of $f_0$ into $K^+K^-$
   channel is subject to large uncertainties. As an educated guess we
   can take the value of $\lambda_n$ for $q^2\bar{q}^2$ states in
   1 GeV mass region from [22]. This yields $\Gamma_{at}\sim 160 eV$
   for $b=0.8 fm$ and $\Gamma_{at}\sim 8 eV$
   for $b=0.2 fm$ as suggested in [5].

   \section{Conclusions}

    We have seen that our model allows on the one hand to include
   kaonium into the realm of the $K\bar{K}$ threshold phenomena, and
   on the other permits to reexamine
       the whole problem from somewhat new point of view. We have
       predicted that the interplay of kaonium and $f_0,a_0$ mesons
       results in drastic behaviour of the amplitude in the vicinity
       of kaonium ground state. The observation of this fascinating
       structure calls for high precision experiments and we are not
       in a position to comment on their feasibility in the
       foreseeable future.

       The effective Hamiltonian we have constructed has provided us
       with interesting insights on the connections between various
       physical quantities: meson positions and widths, threshold
       cusps, isospin breaking effects, kaonium manifestation. We
       have argued that the $f_0-a_0$ mixing definitely influences
       the cross section of reaction (1) in the threshold region.

       In future work it is possible to introduce more sophisticated
       topology of the poles ("shadow poles"), to discuss in more
       detail the $f_0-a_0$ mixing in view of various predictions on
       the corresponding coupling constants. Another point to improve
       is to be more specific concerning the background amplitude
       $T_0$; to this end experimental information on the excitation
       curve around the $K\bar{K}$ threshold is needed. With $T_0$
       really taken into consideration one can use the standard
       machinery [23] to unitarize the complete amplitude.

       Our hope is that we have presented additional arguments in
       favour of precision studies of the $K\bar{K}$ threshold
       region.

       It is a pleasure to thank K.Boreskov, K.Ter-Martirosyan,
       Yu.Kalashnikova and N.Demchuk for useful discussions.

\newpage
\begin{center}
{\large \bf Figure Captions}\\
\end{center}
\vskip 0.5cm

FIG. 1. The diagrams representing the resonance terms in Eq.(18) for the
$\pi \pi$ production amplitude when the invariant mass of the $\pi \pi$
system is close to the energy of the $K^+K^-$ bound state.
\vskip 0.5cm

FIG. 2. The dip-bump structure of the cross section at the vicinity
of the kaonium ground level. This illustration corresponds to the
amplitude given by Eq.(21) with the parameters $V_{at,r}=0.142~MeV~(r=f,a),~
E_r=940~MeV,~ \Gamma_r=214~MeV,  ~ \eta=0.5i,~ T^0=0$.  \vskip 0.5cm

FIG. 3. Shape of the resonance contribution to the amplitude of the
reaction $pd\to ~^3He\pi\pi$. The background amplitude $T^0$ is taken to be
zero. We use the effective Hamiltonian (34) with the following set of
parameters:
$E'_f=E'_a=916~MeV , ~\Gamma_f=\Gamma_a=214~MeV,~D=213~MeV$
and equal $f_0$ and $a_0$ production amplitudes $A_f=A_a$ .
Solid line corresponds
to the similar contributions from $f_0$ and $a_0$ mesons $(\zeta=1)$ and the
dashed line to the $f_0$ dominance $(\zeta=0)$.

   \end{document}